# Comparative analysis of various protocols for high-precision tomography of qudits

Yu.I. Bogdanov*, B.I. Bantysh, N.A. Bogdanova, K. B. Koksharov, V.F. Lukichev

Valiev Institute of Physics and Technology of Russian Academy of Sciences, Moscow, Russia

## ABSTRACT

Quantum tomography is an important tool for obtaining information about the quantum state from experimental data. In this study, we conduct a comparative analysis of various quantum tomography protocols, including protocols based on highly symmetric sets of state vectors, on mutually unbiased bases (MUB), and two-level transformations. Using the universal distribution for the fidelity of quantum state tomography, we estimate the fidelity and efficiency of quantum measurements for qudits of various dimensions as applied to a set of random states uniformly distributed over the Haar measure.

## 1. INTRODUCTION

Quantum states serve as the foundation of modern quantum physics. It turns out that they are also the main resource of new quantum information technologies[1].

Quantum tomography serves as a tool for controlling the technology of preparation and transformation of quantum states. It is aimed at reconstructing quantum states based on statistical results of quantum measurements[2–5].

The present work is aimed at a comparative analysis of the accuracy and efficiency of various protocols of quantum tomography of multilevel quantum systems (qudits). Compared to usual two-level qubits, qudits correspond to the Hilbert spaces of states of higher dimensions. They serve for storing and processing quantum information and can provide a reduction in the complexity of quantum circuits. Qudits can also simplify experimental realization and improve the efficiency of algorithms for quantum computing and quantum communications[6–8].

Currently, several types of state sets are known that define high-quality protocols for quantum tomography of qudits. The most widely used among these sets are symmetric, informationally complete, positive operator-valued measures (SIC-POVM)[9] and mutually unbiased bases (MUB)[10–13].

In our work[14], we have studied a family of tomography protocols produced by solving an optimization problem similar to the Thomson problem. The resulting family includes well-known SIC-POVM and MUB protocols, as well as a number of new highly symmetric sets of quantum states in spaces of dimensions from $s = 2$ to $s = 6$.

The present work has the following structure. In section 2, we give a general idea of quantum measurement protocols and the procedure for extending non-orthogonal unity decomposition to orthogonal one. In the same part, we consider the characteristics of the accuracy of quantum measurements. In section 3, we discuss general properties of the considered protocols. In section 4, we compare their accuracy and efficiency. Finally, in section 5 we present the main conclusions of the study.

---

* bogdanov_yurii@inbox.ru

## 2. QUANTUM TOMOGRAPHY PROTOCOLS

Consider an *s*-dimensional quantum system in the state $|\psi\rangle$. The probability of detecting it in the state $|\phi\rangle$ can be expressed as follows:

$$P = |\langle\phi|\psi\rangle|^2 \tag{1}$$

According to N. Bohr's complementarity principle, various projection measurements form a set of mutually complementary measurements. This set forms the protocol of quantum measurements. It can be compactly represented in the matrix form[15]:

$$M_j = \sum_l X_{jl} c_l, \quad j = 1, 2, ..., m \tag{2}$$

Here $c_l$, $l = 1, 2, ..., s$ are the complex amplitudes of the state $|\psi\rangle$. The protocol describes *m* projections of the quantum state (therefore, matrix *X* contains *m* rows and *s* columns). $M_j$ is the probability amplitude of the *j*-th quantum projection. The probabilities $\lambda_j$ of the corresponding measurements are given by the squared absolute amplitudes:

$$\lambda_j = |M_j|^2 \tag{3}$$

Matrix *X* defines the so-called instrumental matrix of the quantum measurement protocol[15]. It describes the entire set of mutually complementary measurements: the *j*-th row of the matrix *X* defines a bra-vector corresponding to the projective measurement (1).

Consider now a mixed state with the density matrix *ρ*. The probability of registering events corresponding to the *j*-th protocol row is

$$\lambda_j = tr(\Lambda_j \rho) \tag{4}$$

Here $\Lambda_j = X_j^+ X_j$ is the measurement operator, $X_j$ is the *j*-th row of the instrumental matrix *X*.

Consider the following important case:

$$\sum_{j=1}^{m} \Lambda_j = X^+ X = aI, \tag{5}$$

where *a* is some positive constant and *I* is the identity matrix. The protocol meeting condition (5) corresponds to the so-called non-orthogonal unity decomposition. It turns out that by moving to a higher dimensional space, such a protocol can be complemented to traditional von Neumann measurements based on orthogonal unity decomposition[16].

The considered transition to a higher dimensional space is based on the standard procedure of orthogonal (unitary) complement. Let us renormalize the instrumental matrix *X* according to the following rule: $X \to X/\sqrt{a}$. The columns of this new matrix *X* form a set of *s* orthonormal vectors of dimension *m* ($s < m$). We now use that standard orthogonal (unitary) complement procedure. To do this, we add $m - s$ columns to the right. This allows us to expand our rectangular instrumental matrix *X* of dimension $m \times s$ to a square unitary matrix *U* of dimension $m \times m$. Note that the orthogonal complement procedure is multivalued. However, this is not essential for us, since all possible unitary complements correspond to equivalent descriptions of the issue under consideration.

From the physical point of view, we can visually imagine that we are dealing with an *m*-level quantum system (atomic, ionic, molecular, etc.), in which *s* levels (for example, the lowest one) correspond to a qudit, and the remaining $m - s$ levels are auxiliary. The state vector in such an extended *m*-level system is a column of *m* complex probability amplitudes $c_0$, $c_1, ..., c_{m-1}$. At the same time, the preparation of the qudit sets the first *s* amplitudes to $c_0$, $c_1, ..., c_{s-1}$. The rest ones are zero. The implementation of the quantum measurement protocol is reduced to performing a unitary transformation $U$, constructed by the unitary complement procedure with subsequent population readout of all $m$ levels. The physical implementation of this kind of quantum measurement protocols is certainly a very difficult task, especially when the dimension $m$ is large. However, what is important for us is rather the fundamental possibility of this kind of quantum measurements, rather than their technological implementation. Note also that in real experiments, the probabilities corresponding to different rows of the protocol are often measured separately, independently of each other (see for example[5,15]).

We measure the accuracy of pure state reconstruction using fidelity

$$F = |\langle \psi_0 | \psi \rangle|^2 \tag{6}$$

For mixed states the fidelity is

$$F = \left(Tr\sqrt{\rho_0^{1/2} \rho \rho_0^{1/2}}\right)^2 \tag{7}$$

where $\rho_0$ is the theoretical density matrix and $\rho$ is the reconstructed one.

The fidelity losses is a random variable distributed over the generalized $\chi^2$-distribution. To obtain a statistic distribution of losses, we use tomography with the maximum likelihood method[17,18].

The distribution of fidelity losses can also be obtained theoretically using the matrix of complete information $H$[17]. The largest eigenvalue of matrix H corresponds to the normalization of the quantum state. Other $r^2$ zero eigenvalues correspond to the undetectable phases. The remaining $\upsilon_p = (2s-r)r - 1$ non-zero eigenvalues $h_j$ define the vector

$$d_j = \frac{1}{2h_j}, \quad j = 1,...,\upsilon_p. \tag{8}$$

Then the theoretical distribution of fidelity losses is

$$1 - F = \sum_{j=1}^{\upsilon_p} d_j \xi_j^2, \tag{9}$$

where $\xi_j \sim N(0,1)$ are independent standard normal random variables. It is convenient to introduce the losses function that is asymptotically independent of the sample size[18]:

$$L = N\langle 1 - F \rangle \tag{10}$$

The efficiency of the quantum protocol is the ratio between the average fidelity losses and its theoretical minimum:

$$\eta = \frac{\langle 1-F \rangle_{\min}}{\langle 1-F \rangle} = \frac{\upsilon_p^2}{4L(s-1)} \tag{11}$$

Note that, in the general case, for a space of dimension $s$, for the protocols forming the decomposition of unity, the theoretical accuracy limit is $L_{\min}^{theor} = s - 1$ [18].

## 3. BRIEF DESCRIPTION OF THE CONSIDERED PROTOCOLS OF QUANTUM TOMOGRAPHY

Along with the highly symmetric sets of projection states in spaces of various dimensions proposed in our work[14], we considered protocols based on sets of two-level qudits transformation[19] as well as protocols based on projection measurements in mutually unbiased based (MUB).

Two-level protocol is based on performing some transformation between two levels of a qudit with subsequent readout of these levels. The particular transformations are $U_1 = \frac{1}{\sqrt{2}}\begin{pmatrix} 1 & 1 \\ 1 & -1 \end{pmatrix}$ and $U_2 = \frac{1}{\sqrt{2}}\begin{pmatrix} 1 & 1 \\ i & -i \end{pmatrix}$. Their columns are the eigenvectors of $\sigma_x$ and $\sigma_y$ Pauli matrices respectively. The number of distinct pairs of levels is $C_s^2 = s(s-1)/2$. Each two-level unitary generates two rows in the protocol instrumental matrix $X$, while the matrices $U_1$ and $U_2$ generate 4 rows together. Thus, there are $4C_s^2 = 2s(s-1)$ rows. In addition, the populations of all $s$ are measured, which generates $s$ more protocol rows. Thus, the full number protocol rows is: $m = 2s(s-1) + s = s(2s-1)$

MUB protocols are considered in spaces of dimensions $s = 2, 3, 4, 5, 8$. Let the vectors $|e_j\rangle$ and $|f_j\rangle$ be taken from the sets forming two different orthogonal bases. These bases are called mutually unbiased if the following relation are satisfied[10–13]:

$$|\langle e_i | f_j \rangle|^2 = \frac{1}{s}, \; i, j = 0, ..., s-1. \tag{12}$$

It is known that informationally complete sets of mutually unbiased bases exist in spaces whose dimension is equal to the power of a prime number. At the same time, the existence of these sets in spaces of arbitrary dimensions has not yet been proven. According to the results of modern numerical experiments, the complete set of mutually unbiased bases is probably already missing in the space of dimension six. It is known that the maximum number of mutually unbiased bases in space of dimension $s$ is $s + 1$, provided that $s$ is an integer power of a prime number. Under such conditions, by MUB measurement one means the measurements in all $s + 1$ mutually unbiased bases. This provides informational completeness, which in turn makes it possible to reconstruct an arbitrary state (pure or mixed). The number of MUB protocol rows is $m = s(s+1)$.

Let us write down explicitly the elements of these bases for systems of dimensions 2, 3 and 4[13]. We represent each basis of the $s$-dimensional MUB protocol in the form of unitary matrices $U_j$ ($j = 0, ..., s$) of dimension $s \times s$, in which column correspond to the basis vectors. Note that the rows of the instrumental matrix $X$ correspond to the rows of the matrices $U_j^+$.

For $s = 2$ (qubit)

$$U_0 = \begin{pmatrix} 1 & 0 \\ 0 & 1 \end{pmatrix}, U_1 = \frac{1}{\sqrt{2}}\begin{pmatrix} 1 & 1 \\ 1 & -1 \end{pmatrix}, U_2 = \frac{1}{\sqrt{2}}\begin{pmatrix} 1 & 1 \\ i & -i \end{pmatrix} \tag{13}$$

These bases correspond to the measurement of the observed Pauli $\sigma_z$, $\sigma_x$ and $\sigma_y$, respectively. For $s = 3$ (qutrit):

$$U_0 = \begin{pmatrix} 1 & 0 & 0 \\ 0 & 1 & 0 \\ 0 & 0 & 1 \end{pmatrix}, U_1 = \frac{1}{\sqrt{3}}\begin{pmatrix} 1 & 1 & 1 \\ 1 & \omega_3 & \omega_3^2 \\ 1 & \omega_3^2 & \omega_3 \end{pmatrix},$$

$$U_2 = \frac{1}{\sqrt{3}}\begin{pmatrix} 1 & 1 & 1 \\ \omega_3 & \omega_3^2 & 1 \\ \omega_3 & 1 & \omega_3^2 \end{pmatrix}, U_3 = \frac{1}{\sqrt{3}}\begin{pmatrix} 1 & 1 & 1 \\ \omega_3^2 & \omega_3 & 1 \\ \omega_3^2 & 1 & \omega_3 \end{pmatrix} \tag{14}$$

where $\omega_3 = \exp(2\pi i/3)$. For $s = 4$ (ququart)

$$U_0 = \begin{pmatrix} 1 & 0 & 0 & 0 \\ 0 & 1 & 0 & 0 \\ 0 & 0 & 1 & 0 \\ 0 & 0 & 0 & 1 \end{pmatrix}, U_1 = \frac{1}{2}\begin{pmatrix} 1 & 1 & 1 & 1 \\ 1 & 1 & -1 & -1 \\ 1 & -1 & -1 & 1 \\ 1 & -1 & 1 & -1 \end{pmatrix}, U_2 = \frac{1}{2}\begin{pmatrix} 1 & 1 & 1 & 1 \\ -1 & -1 & 1 & 1 \\ -i & i & i & -i \\ -i & i & -i & i \end{pmatrix}$$

$$U_3 = \frac{1}{2}\begin{pmatrix} 1 & 1 & 1 & 1 \\ -i & -i & i & i \\ -i & i & i & -i \\ -1 & 1 & -1 & 1 \end{pmatrix}, U_4 = \frac{1}{2}\begin{pmatrix} 1 & 1 & 1 & 1 \\ -i & -i & i & i \\ -1 & 1 & -1 & 1 \\ -i & i & i & -i \end{pmatrix}$$
(15)

## 4. ANALYSIS RESULTS

Figures 1-4 show the results of the analysis of the accuracy and efficiency of tomography of pure states of qutrites ($s = 3$) for various protocols. These included 2-level ($m = 15$), MUB ($m = 12$) and symmetric ($m = 30$) protocols.

The sample was made up of 10 thousand random pure states of qutrit from a uniform distribution according to the Haar measure. The average fidelity loss was $L_{mean}^{MUB} = 2.20530 \pm 0.00091$ for the MUB protocol and $L_{mean}^{2-level} = 2.2146 \pm 0.0015$ for the 2-level protocol.

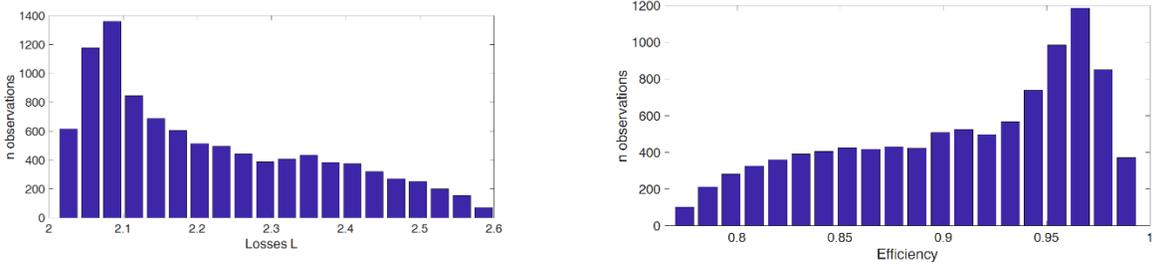

Figure 1. Losses of fidelity (left), efficiency (right); $s = 3$, 2-level, $m = 15$.

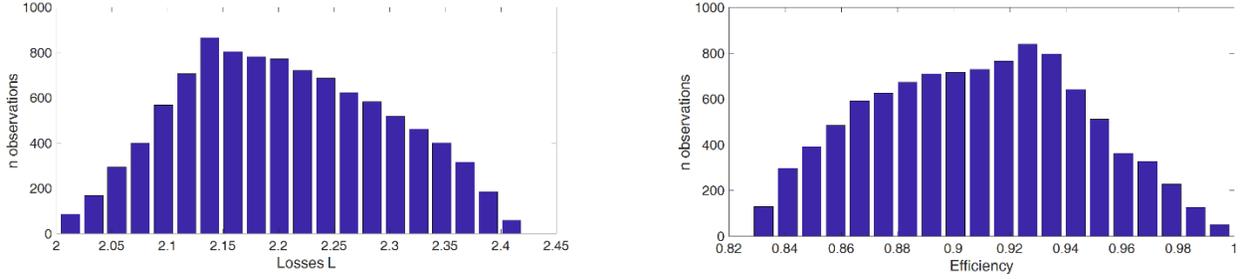

Figure 2. Losses of fidelity (left), efficiency (right); $s = 3$, MUB, $m = 12$.

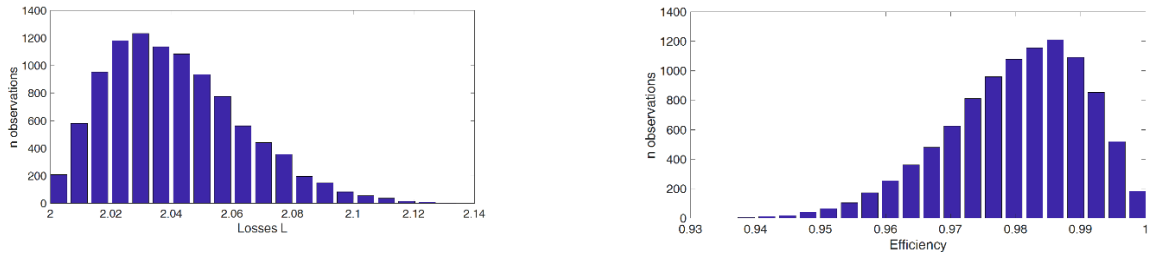

Figure 3. Losses of fidelity (left), efficiency (right); $s = 3$, symmetric, $m = 30$.

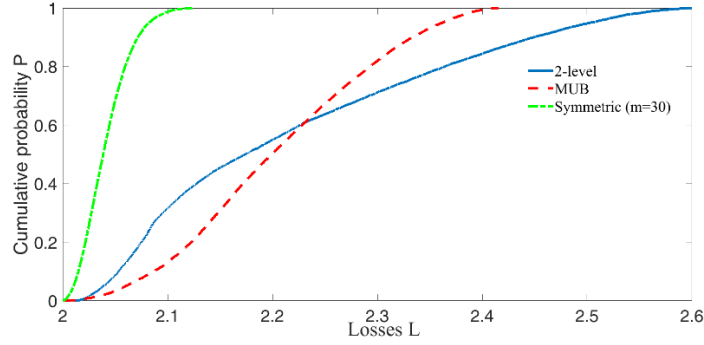

Figure 4. Cumulative losses of fidelity in tomography of qutrits; $s = 3$.

Thus, the MUB protocol has slightly lower average losses compared to the 2-level protocol. This difference, taking into account the presented values of the mean square errors, is statistically significant. At the same time, the 2-level protocol has a lower median value compared to the MUB protocol: $L_{50\%}^{2-level} \approx 2.17$ against $L_{50\%}^{MUB} \approx 2.20$.

Figure 4 shows that the cumulative curves for 2-level and MUB protocols intersect at a point that approximately corresponds to the 61-st percentile. Up to this point, the 2-level protocol loss curve lies to the left of the MUB protocol curve. This means that in approximately 61 percent of cases, the 2-level protocol is superior in accuracy to the MUB protocol (and this despite the fact that "on average", as mentioned above, the 2-level protocol is less accurate).

In Figure 4, we can also see that the symmetric 30-row protocol significantly outperforms both the 2-level and the MUB protocol. This is because the corresponding cumulative curve lies significantly to the left of the other two. At the same time, the average loss of accuracy for a symmetric protocol with 30 rows was $L_{mean}^{Symmetric,m=30} = 2.04134 \pm 0.00022$. Thus, the protocol under consideration is very close to the theoretical limit $L_{min}^{theor} = 2$. However, the implementation of the considered high-precision protocol with 30 rows is much more complicated than the implementation of the other two. This requires considering 30 measurement operators instead of 12 operators for the MUB protocol and 15 operators for the 2-level protocol.

In the present work, various highly symmetric protocols for qudits of dimensions' $s = 3, 4, 5, 6$ have been investigated. The protocols under consideration are reduced to POVM or are close to such. In addition, all the protocols under consideration have the property of tomographic completeness[17,18].

Figure 5 shows graphs of the average loss of accuracy for pure states, uniformly distributed in the Haar measure, for symmetric protocols for various dimensions' $s = 3, 4, 5, 6$ and for various number $m$ of protocol rows. Each point on the graphs corresponds to a numerical experiment with 10 thousand random pure states.

For dimensions $s = 3, 4, 5$ at $m = s(s+1)$, the symmetric protocols that we are considering, and that are constructed there[14], coincide with the corresponding MUB protocols. This property, however, does not hold for $s = 6$. We note that for the case $s = 6$, no MUB protocols are known at all. We present the values of fidelity losses of MUB protocols for the dimensions of $s = 3, 4, 5$. For $s = 3$ and $m = 12$, we get $L_{mean}^{MUB,s=3} = 2.20440 \pm 0.00091$, which, within the limits of statistical errors, is consistent with the value provided above. Similarly, for $s = 4$ and $m = 20$ we have $L_{mean}^{MUB,s=4} = 3.27362 \pm 0.00082$, and for $s = 5$ and $m = 30$ we obtain $L_{mean}^{MUB,s=5} = 4.3245 \pm 0.0011$. Note that in the case of $s = 5$ 2-level protocol contains $m = 45$ rows and has losses that turn out to be higher than those of the corresponding MUB protocol.

For a complete picture, let us also present the results of similar numerical experiments for the MUB protocols at $s = 2$ and $s = 8$. For $s = 2$ and $m = 6$ we have $L_{mean}^{MUB,s=2} = 1.08371 \pm 0.00037$ (herewith $1 \leq L \leq 1.125$). Note that in this particular case the protocols of all three considered types coincide (2-level, MUB and highly symmetrical). Similar calculations for $s = 8$ and $m = 72$ give $L_{mean}^{MUB,s=8} = 7.39304 \pm 0.00062$. In this case, the observed values of the fidelity losses are again very close to the theoretical limit $L_{min}^{theor} = 7$. Note that for a 2-level protocol with $s = 8$, we can get $L_{mean}^{2-level,s=8} = 8.1848 \pm 0.0049$. Apparently, in the case under consideration, the 2-level protocol is noticeably worse than the MUB protocol, and at the same time requires a larger number of measurement rows ($m = 120$ for 2-level protocol instead of $m = 72$ for MUB protocol).

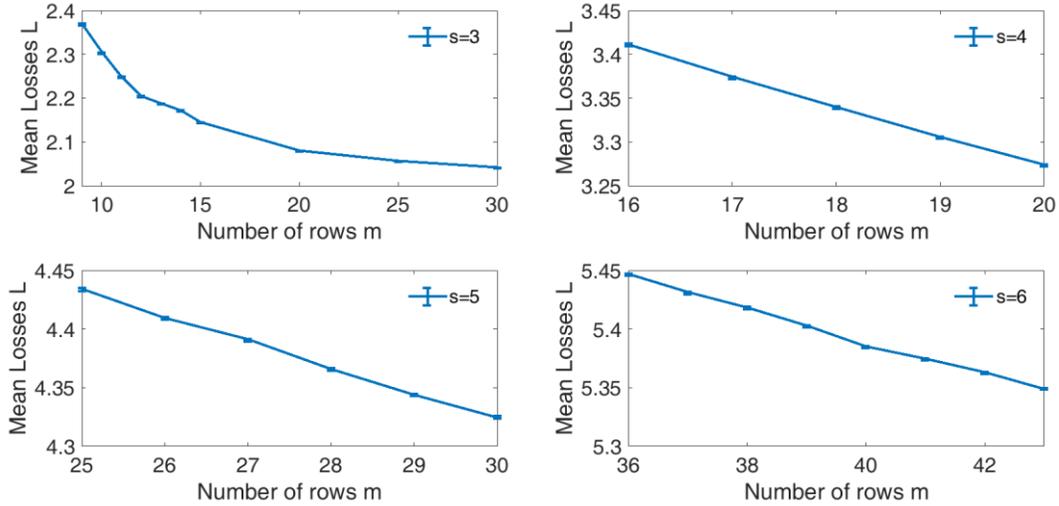

Figure 5. Plots of average fidelity losses of symmetric protocols for different dimensions and different number $m$ of protocol rows; $s = 3, 4, 5, 6$.

Note also that for 2-level protocol with $s = 16$ and $m = 296$, you can get $L_{mean}^{2-level,s=16} = 17.8667 \pm 0.0080$. The resulting value gives a good accuracy, since in the case under consideration $L_{min}^{theor} = 15$.

Finally, consider examples of mixed states of full rank ($r=s$), distributed according to the Hilbert-Schmidt measure (which is an analogue of the Haar measure for pure states[20]). Figure 6 below shows the case $s = 8$.

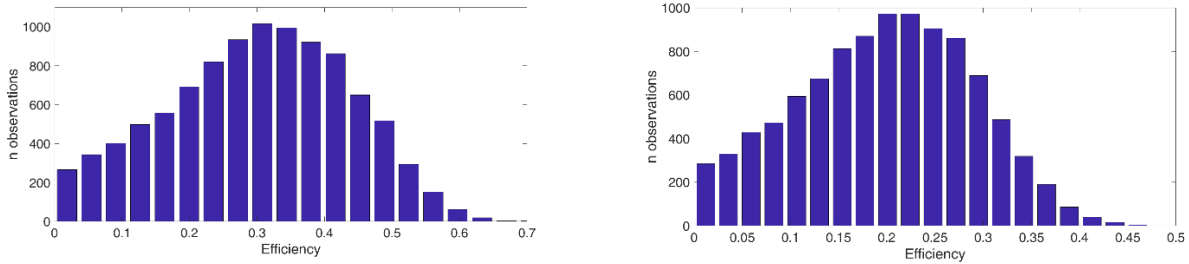

Figure 6. Distribution of efficiency for mixed states of full rank for $s = 8$. On the left is the MUB protocol, on the right is 2-level.

10 thousand experiments were performed for each protocol. The average values of efficiency for the MUB protocol and 2-level protocol were, respectively: $Eff_{mean}^{MUB,s=8,r=8} = 0.3000 \pm 0.0013$, $Eff_{mean}^{2-level,s=8,r=8} = 0.19753 \pm 0.00091$.

We see that in the case under consideration, the MUB protocol is, on average, one and half times more efficient than the 2-level protocol.

The peculiarity of the analysis of mixed states is that for some states the efficiency values turn out to be small (close to zero), which corresponds to high values of fidelity losses. Here, in contrast to pure states, it is impossible to specify an upper bound for the fidelity losses (states with arbitrarily high values of fidelity losses are possible).

# 5. CONCLUSIONS

– A comparative analysis of various protocols of quantum tomography of qudits, including protocols based on highly symmetric sets of state vectors, on mutually unbiased bases, as well as on 2-level transformations, is carried out.

– Based on the complete information matrix and the universal distribution for the accuracy of statistical reconstruction of quantum states, the accuracy and efficiency of quantum measurements for qudits of various dimensions is analyzed with respect to a set of random states uniformly distributed in the Haar measure.

– For a number of quantum measurement protocols that have found wide scientific and practical use, fundamental characteristics of accuracy and efficiency have been obtained.

– The results obtained are of great importance for the development, monitoring and prediction of the characteristics of high-precision control of multilevel quantum information systems.

# 6. ACKNOWLEDGMENTS

The investigation was supported by the Program no. FFNN-2022-0016 of the Ministry of Science and Higher Education of Russia for Valiev Institute of Physics and Technology of RAS, and by the Foundation for the Advancement of Theoretical Physics and Mathematics BASIS (project no. 20-1-1-34-1).